\RequirePackage[mathlines]{lineno} 
\documentclass[prd,twocolumn,showpacs,amsmath,amssymb,aps]{revtex4}
\usepackage{overpic,graphicx}
\usepackage{dcolumn}
\usepackage{bm}
\usepackage{rotating}
\usepackage{times}
\usepackage{amsfonts}
\usepackage{amsthm}
\usepackage{latexsym}
\usepackage{color}
\usepackage{url}
\usepackage{subfigure}

\begin{document}
\normalsize
\parskip=5pt plus 1pt minus 1pt

\title{ \boldmath Measurements of the absolute branching fractions for
$D_{s}^{+}\rightarrow\eta e^{+}\nu_{e}$ and $D_{s}^{+}\rightarrow\eta^{\prime} e^{+}\nu_{e}$
}
\author{\small{
M.~Ablikim$^{1}$, M.~N.~Achasov$^{9,e}$, S. ~Ahmed$^{14}$, X.~C.~Ai$^{1}$, O.~Albayrak$^{5}$, M.~Albrecht$^{4}$, D.~J.~Ambrose$^{45}$, A.~Amoroso$^{50A,50C}$, F.~F.~An$^{1}$, Q.~An$^{47,a}$
% ...
\\
\vspace{0.2cm}
(BESIII Collaboration)\\
}}
\vspace{0.2cm} 
\affiliation{$^{1}$ Institute of High Energy Physics, Beijing 100049, People's Republic of China}
% ...
\affiliation{$^{4}$ Bochum Ruhr-University, D-44780 Bochum, Germany}
\affiliation{$^{5}$ Carnegie Mellon University, Pittsburgh, Pennsylvania 15213, USA}
% ...
\affiliation{$^{9}$ G.I. Budker Institute of Nuclear Physics SB RAS (BINP), Novosibirsk 630090,
Russia}
% ...
\affiliation{$^{14}$ Helmholtz Institute Mainz, Johann-Joachim-Becher-Weg 45, D-55099 Mainz, Germany}
% ...
\affiliation{$^{45}$ University of Rochester, Rochester, New York 14627, USA}
% ...
\affiliation{$^{47}$ University of Science and Technology of China, Hefei 230026, People's Republic of
China}
% ...
\affiliation{$^{50}$ (A)University of Turin, I-10125, Turin, Italy; (B)University of Eastern Piedmont,
I-15121, Alessandria, Italy; (C)INFN, I-10125, Turin, Italy}
% ...
\vspace{0.2cm}
\affiliation{$^{a}$ Also at State Key Laboratory of Particle Detection and Electronics, Beijing 100049,
Hefei 230026, People's Republic of China}
\affiliation{$^{b}$ Also at Bogazici University, 34342 Istanbul, Turkey}
\affiliation{$^{c}$ Also at the Moscow Institute of Physics and Technology, Moscow 141700, Russia}
\affiliation{$^{d}$ Also at the Functional Electronics Laboratory, Tomsk State University, Tomsk,
634050, Russia}
\affiliation{$^{e}$ Also at the Novosibirsk State University, Novosibirsk, 630090, Russia}
\affiliation{$^{f}$ Also at the NRC "Kurchatov Institute, PNPI, 188300, Gatchina, Russia}
\affiliation{$^{g}$ Also at University of Texas at Dallas, Richardson, Texas 75083, USA}
\affiliation{$^{h}$ Also at Istanbul Arel University, 34295 Istanbul, Turkey}
\affiliation{$^{i}$ Also at Goethe University Frankfurt, 60323 Frankfurt am Main, Germany}
\affiliation{$^{j}$ Also at Institute of Nuclear and Particle Physics, Shanghai Key Laboratory for
Particle Physics and Cosmology, Shanghai 200240, People's Republic of China}

\vspace{0.4cm}

\begin{abstract}
By analyzing 482 pb$^{-1}$ of $e^+e^-$ collision data collected
at $\sqrt s=4.009$ GeV with the BESIII detector at the BEPCII
collider, we measure the absolute branching fractions for the semileptonic
decays $D_{s}^{+}\to\eta e^{+}\nu_{e}$ and $D_{s}^{+}\to
\eta'e^{+}\nu_{e}$
to be ${B}(D_{s}^{+}\rightarrow\eta e^{+}\nu_{e})=(2.30\pm0.31\pm0.08)$\% and
${B}(D_{s}^{+}\rightarrow\eta'e^{+} \nu_{e}) = (0.93\pm0.30\pm0.05)$\%, respectively,
and their ratio $\frac{{B}(D_{s}^{+}\rightarrow\eta'e^{+}\nu_{e})}
{{B}(D_{s}^{+}\rightarrow\eta e^{+}\nu_{e})}=0.40\pm0.14\pm0.02$,
where the first uncertainties are statistical and the second ones are systematic.
The results are in good agreement with previous measurements within uncertainties;
they can be used to determine the $\eta-\eta'$ mixing angle and improve upon the
$D_s^+$ semileptonic branching ratio precision.
\end{abstract}
\maketitle

\section{\boldmath Introduction}
The semileptonic decays $D^+_s \to \eta e^+\nu_e$ and
$D^+_s \to \eta'e^+\nu_e$
are important channels for the study of
heavy quark decays and light meson spectroscopy.
The inclusive semileptonic
decay widths of the mesons $D^0$, $D^+$ and $D^+_s$
should be equal, up to $SU(3)$ symmetry breaking and non-factorizable components~\cite{OPE}.
The measured inclusive semileptonic decay widths of $D^0$ and $D^+$
mesons are proven to be consistent with each other. However, they are
larger than that of $D^+_s$ mesons by 20\%~\cite{prd81},
more than $3\sigma$ of the experimental uncertainties.
The updated Isgur-Scora-Grinstein-Wise form factor model (ISGW2)~\cite{ISGW2}
predicts a difference between
the $D$ and $D^+_s$ inclusive semileptonic rates,
as the spectator quark masses $m_{u}$ and $m_s$ differ on the scale of
the daughter quark mass $m_s$ in the Cabibbo favored semileptonic transition.
Up to now, the exclusive semileptonic decays
of $D^0$ and $D^+$ mesons have been well studied experimentally~\cite{pdg2014}.
Therefore, measurements of the $D^+_s$ exclusive semileptonic decay rates
will provide helpful information to understand this difference.
In addition, it is well known that the states $\eta$ and $\eta'$ are considered as
candidates for mixing with gluonic components.
The exclusive semileptonic decays
$D^+_s \to \eta e^+\nu_e$ and $D^+_s \to \eta'e^+\nu_e$
probe the $s\bar s$ components of $\eta$ and $\eta'$
and thus are sensitive to the $\eta-\eta'$ mixing angle~\cite{MixAngle}.
Therefore, measurements of these decay rates
can constrain the physics related to the mixing with the gluonic components~\cite{Glueball}.

The CLEO Collaboration measured the ratio between
the branching fractions for $D_{s}^{+}\rightarrow\eta' e^{+}\nu_{e}$
and $D_{s}^{+}\rightarrow\eta e^{+}\nu_{e}$ to be
$\frac{B(D^+_s\to\eta' e^+\nu_e)}{B(D^+_s\to \eta e^+\nu_e)} = 0.35\pm0.09\pm0.07,$
by analyzing a data sample of 3.11 fb$^{-1}$
taken at the center-of-mass
energy $\sqrt s$ at
$\varUpsilon(4S)$ in 1995~\cite{CLEORatio}, and the two individual branching fractions
to be $B(D_{s}^{+}\rightarrow\eta e^{+}\nu_{e})$ = ($2.48\pm0.29\pm0.13$)\%
and $B(D_{s}^{+}\rightarrow\eta'e^{+}\nu_{e})$=($0.91\pm0.33\pm0.05$)\% using
a data sample of 310 pb$^{-1}$ collected with the CLEO-c detector
at $\sqrt s=4.17$ GeV in 2009~\cite{prd_80_052007}.
Recently, these two branching fractions were measured to be
$B(D_{s}^{+}\rightarrow\eta e^{+}\nu_{e})$ = ($2.28\pm0.14\pm0.20$)\%
and $B(D_{s}^{+}\rightarrow\eta'e^{+}\nu_{e})$=($0.68\pm0.15\pm0.06$)\%,
by using a data sample of 586 pb$^{-1}$ collected
at $\sqrt s=4.17$ GeV with the CLEO-c detector~\cite{arx_1505_04205}.
In this paper, we report measurements of the absolute branching
fractions for $D_{s}^{+}\rightarrow\eta e^{+}\nu_{e}$ and
$D_{s}^{+}\rightarrow\eta'e^{+}\nu_{e}$ at the BESIII experiment.

\section{\boldmath Detector and monte carlo}
This analysis presented in this paper is carried out using a data sample of 482~pb$^{-1}$~\cite{data}
collected at $\sqrt s=4.009$ GeV~with the BESIII detector.

BESIII is a cylindrical spectrometer that is
composed of a Helium-gas based main drift chamber
(MDC), a plastic scintillator time-of-flight (TOF) system,
a CsI (Tl) electromagnetic calorimeter (EMC), a
superconducting solenoid providing a 1.0 T magnetic field
and a muon counter in the iron flux return yoke of the magnet. The charged particle
momentum resolution is 0.5\% at a transverse momentum of
1 GeV/$c$, and the photon energy resolution is 2.5\% at
an energy of 1 GeV. Particle identification (PID) system combines the
ionization energy loss ($dE/dx$) in MDC, the TOF and
EMC information to identify particle types.
More details about BESIII are described in Ref.~\cite{bes3}.

A GEANT4-based ~\cite{GEANT4} Monte Carlo (MC)
simulation software, which includes the geometric description of
the BESIII detector and its response, is used to determine the
detection effciency and estimate background contributions.
The simulation is implemented with KKMC~\cite{KKMC}, EVTGEN~\cite{EvtGen,EvtGen1} and PHOTOS~\cite{PHOTOS}
and includes the effects of Initial State Radiation (ISR) and Final State Radiation (FSR).
A generic MC sample (called `inclusive MC sample' hereafter)
corresponding to an equivalent integrated luminosity of 11~fb$^{-1}$
includes open charm production, ISR production of low-mass vector charmonium states,
continuum light quark production, $\psi(4040)$ decays and QED events.
The known decay modes of the charmonium states are produced by EVTGEN with the branching fractions
being set to world average values~\cite{pdg2014}, and the remaining,
unknown, ones are simulated by LUNDCHARM~\cite{LundCharm}. The semileptonic decays
are generated with the ISGW2 form factor model~\cite{ISGW2}.
                                                                  
\section{\boldmath Singly Tagged $D_{s}^{-}$ events}
At $\sqrt s= 4.009\;\text{GeV}$, the $\psi(4040)$ resonance is produced in
electron-positron ($e^{+}e^{-}$) annihilation. The $\psi(4040)$ lies just above
the charm-strange meson pair $D_{s}^{+}D_{s}^{-}$ production threshold and decays into
a $D_{s}^{+}D_{s}^{-}$ pair in a clean way, with no additional
particles in the final state. If one $D_{s}^{-}$ meson is fully reconstructed
(called a singly tagged (ST) $D_{s}^{-}$ event),
the presence of a $D_{s}^{+}$ meson on the recoil side can be inferred.
In this analysis, the ST $D_{s}^{-}$
mesons are reconstructed in ten hadronic decay modes:
$K^+K^-\pi^{-}$,
$\phi\rho^-$($\phi\to K^+K^-,\rho^-\to\pi^{0}\pi^{-}$),
$K^{0}_{S}K^{+}\pi^{-}\pi^{-}$, $K^{0}_{S}K^{-}\pi^{+}\pi^{-}$,
$K^{0}_{S}K^{-}$, $\pi^+\pi^-\pi^-$, $\eta\pi^{-}$ ($\eta\to\gamma\gamma$),
$\eta'\pi^{-}$($\eta'\to\eta\pi^+\pi^-$,$\eta\to\gamma\gamma$),
$\eta'\pi^{-}$($\eta'\to\gamma\rho^0$), $\eta\rho^-$($\eta\to\gamma\gamma$).
Throughout the paper, charge conjugation is implied, and the ST
modes are selected separately according to their charge.

We require that all the charged tracks are well reconstructed
in the MDC with good helix fits, and their polar angles in the MDC must satisfy $|\cos\theta|<0.93$.
For each charged track, save those from $K^{0}_{S}$ decays,
the point of closest approach to the $e^+e^-$ interaction point (IP) must be within
$\pm$10 cm along the beam direction and within 1 cm in the plane perpendicular to the beam direction.
For charged particle identification, the combined confidence levels
for the pion and kaon hypotheses, $CL_{\pi}$ and $CL_{K}$,
are calculated using the $dE/dx$ and TOF information.
A charged track satisfying $CL_{\pi}>0$ and $CL_{\pi}>CL_{K}$
($CL_{K}>0$ and $CL_{K}>CL_{\pi}$) is identified as a pion (kaon).

The $K^{0}_{S}$ candidates are reconstructed from pairs of oppositely charged tracks.
For these two tracks,
the point of the closest approach
to the IP must be within $\pm$20 cm along the beam direction.
The two oppositely charged tracks are assigned as $\pi^+\pi^-$ without PID.
The $\pi^+\pi^-$ invariant mass is required to satisfy
$0.487<M(\pi^+\pi^-)<0.511$ GeV/$c^2$. The two
tracks are constrained to originate from a common decay vertex,
which is required to have a positive separation from the IP with respect to the
$K^{0}_{S}$ flight direction.

Photon candidates are reconstructed from clusters in the EMC.  The energy deposited in
nearby TOF counters is included to improve the reconstruction
efficiency and energy resolution. Showers must have minimum energy of 25 MeV
in the barrel region ($|\cos\theta|<0.80$) or 50 MeV in the end cap region
($0.86<|\cos\theta|<0.92$).
To suppress electronic noise and clusters unrelated to the event,
the EMC cluster time is required to be within [0, 700]~ns after the
event start time.
The angle between the photon candidates and the closest charged track
is required to be greater than $10^\circ$
to suppress split-off showers or bremsstrahlung generated by charged particles.

The $\pi^0$ and $\eta$ candidates are reconstructed from photon pairs.
We require that the $\gamma\gamma$ invariant mass satisfies $0.115<M(\gamma\gamma)<0.150$ GeV/$c^2$
for $\pi^0$ candidates, and $0.510<M(\gamma\gamma)<0.570$ GeV/$c^2$
for $\eta$ candidates.
To improve the mass resolution, a mass-constrained fit to the nominal mass of
$\pi^0$ or $\eta$~\cite{pdg2014} is applied to the photon pairs.

For $\phi$ and $\rho^-$ candidates, the invariant mass is required to satisfy
$1.005<M(K^+K^-)<1.040$ GeV/$c^2$ and $0.570<M(\pi^0\pi^-)<0.970$ GeV/$c^2$, respectively.
For $\eta'$ candidates, the invariant mass must satisfy
$0.943<M(\eta'_{\eta\pi^+\pi^-})<0.973$ GeV/$c^2$ or
$0.932<M(\eta'_{\gamma\rho^0})<0.980$ GeV/$c^2$, we additionally require
$0.570<M(\pi^+\pi^-)<0.970$ GeV/$c^2$ for $\eta'_{\gamma\rho^0}$ candidates
to reduce contributions from combinatorial background.

The ST $D_{s}^{-}$ meson is identified
using the energy difference $\Delta E \equiv E_{\rm ST}-E_{\rm beam}$ and the beam energy constrained mass
$M_{\rm BC} \equiv \sqrt{E^2_{\rm beam}-|\overrightarrow{p}_{\rm ST}|^{2}}$,
where $E_{\rm ST} = \Sigma_i E_i$ and $|\overrightarrow{p}_{\rm ST}|=|\Sigma_i \overrightarrow{p}_i|$
are the total energy and momentum of all the final state particles of the ST system,
and $E_{\rm beam}$ is the beam energy.
In order to improve the ratio of signal to background, the
$\Delta E$ is required to fall in a ($-3\sigma$, $3\sigma$)
window around the peak of the $\Delta E$ distribution, where $\sigma$ is
the standard deviation of the $\Delta E$ distribution.
For each ST mode, if more than one combination satisfies the criteria in an event,
only the combination with the minimum $|\Delta E|$ is retained.

\begin{figure}[htbp]
\includegraphics[width=8.6cm,height=15.6cm]{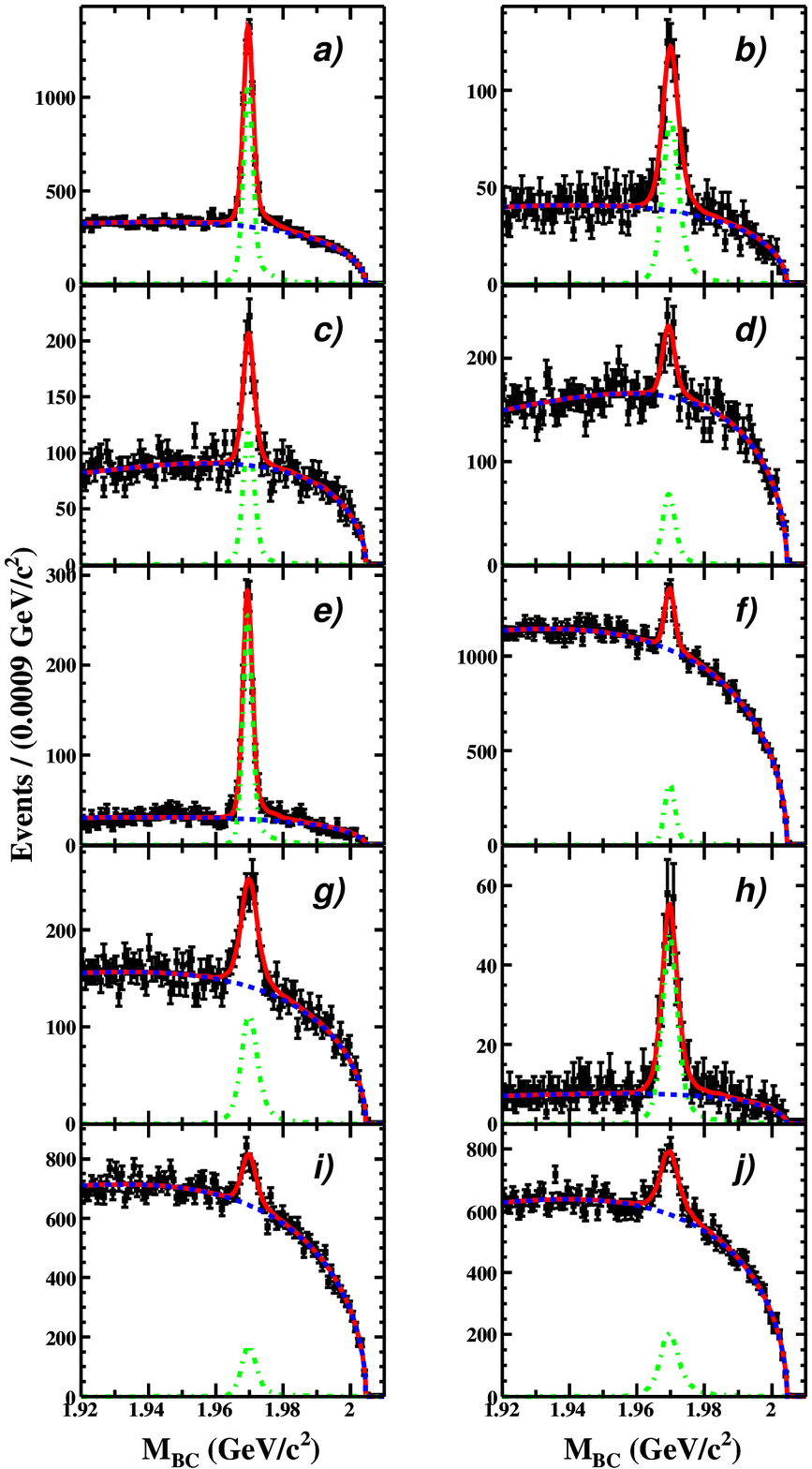}
 \caption{ \label{fig:Stag_Mbc}
  Results of the fits to the $M_{\rm BC}$ distributions of the
  ST $D_{s}^{-}$ modes
  (a) $K^+K^-\pi^-$, (b) $\phi\rho^-,\phi\to K^+K^-$,
  (c) $K^{0}_{S}K^+\pi^-\pi^-$, (d) $K^{0}_{S}K^-\pi^+\pi^-$,
  (e) $K^{0}_{S}K^-$, (f) $\pi^+\pi^-\pi^-$, (g) $\eta\pi^-,\eta\to\gamma\gamma$,
  (h) $\eta'\pi^-,\eta'\to\eta\pi^+\pi^-$, (i) $\eta'\pi^-,\eta'\to\gamma\rho^0$,
  (j) $\eta\rho^-,\eta\to\gamma\gamma$. In each plot, the dots
  with error bars are from data, the red solid curve represents the
  total fit to the data,
  the blue dashed curve describes the ARGUS background, and the green dotted curve
  denotes the signal shape.}
\end{figure}

To determine the number of ST $D_{s}^{-}$ mesons, we perform a fit to
the $M_{\rm BC}$ spectra of the accepted combinations.
In the fits, we use the MC simulated signal shape convoluted with
a Gaussian function to represent the signal shape
and an ARGUS function~\cite{argus} to describe the background,
which is expected to be a smooth distribution in $M_{\rm BC}$.
The fits to the $M_{\rm BC}$ spectra are shown in Fig.~\ref{fig:Stag_Mbc}.
The events in the $M_{\rm BC}$ signal region, which is defined to be within
a ($-4\sigma$, $5\sigma$) window around
the peak of the $M_{\rm BC}$ distribution, are kept for
further analysis.
The numbers of the ST $D_{s}^{-}$ mesons are obtained by integrating
the $D_{s}^{-}$ signal over the $M_{\rm BC}$ signal region.
We estimate the efficiency of reconstructing the ST $D_{s}^{-}$ mesons
(ST efficiency $\epsilon_{D^{-}_{s}}^{\rm ST}$) by analyzing the inclusive $D_{s}^{+}D_{s}^{-}$ MC sample.
The requirements on $\Delta E$ and $M_{\rm BC}$,
the numbers of the ST $D_{s}^{-}$ mesons and the ST
efficiencies are summarized in Tab.~\ref{tab:SingleTag}.
The total number ($N_{\rm ST}^{\rm tot}$) of the ST $D_{s}^{-}$ mesons is $13157\pm240$.
\begin{table*}[hbtp]
\caption{\label{tab:SingleTag}
Summary of the requirements on $\Delta E$ and $M_{\rm BC}$,
the numbers of the ST $D_{s}^{-}$ ($N_{\rm ST}$) in data
and the ST efficiencies ($\epsilon_{D^{-}_{s}}^{\rm ST}$) which do not include the branching fractions for
daughter particles of $\pi^0$, $K^{0}_{S}$, $\eta$ and $\eta'$.
Charge conjugation is implied, and the uncertainties are
statistical only.
}
\renewcommand\arraystretch{1.1}
\begin{tabular}{lcccc} \hline\hline
Tag Mode & $\Delta E$ (GeV) & $M_{\rm BC}$ (GeV/$c^2$) & $N_{\rm ST}$ & $\epsilon_{D^{-}_{s}}^{\rm ST}$ (\%) \\ \hline
$K^+K^-\pi^-$                      &$(-0.020,0.017)$&$(1.9635,1.9772)$&$4863\pm95 $&$38.92\pm0.08$ \\
$\phi(K^+K^-)\rho^-$                &$(-0.036,0.023)$&$(1.9603,1.9821)$&$616 \pm39 $&$10.05\pm0.07$ \\
$K^{0}_{S}K^+\pi^-\pi^-$           &$(-0.018,0.014)$&$(1.9632,1.9778)$&$601 \pm40 $&$23.17\pm0.16$ \\
$K^{0}_{S}K^-\pi^+\pi^-$           &$(-0.016,0.012)$&$(1.9622,1.9777)$&$388 \pm52 $&$21.98\pm0.21$ \\
$K^{0}_{S}K^-$                     &$(-0.019,0.020)$&$(1.9640,1.9761)$&$1078\pm38 $&$44.96\pm0.20$ \\
$\pi^+\pi^-\pi^-$                  &$(-0.026,0.022)$&$(1.9634,1.9770)$&$1525\pm116$&$51.83\pm0.14$ \\
$\eta(\gamma\gamma)\pi^-$          &$(-0.052,0.058)$&$(1.9598,1.9824)$&$840 \pm56 $&$47.58\pm0.24$ \\
$\eta'(\eta\pi^+\pi^-)\pi^-$       &$(-0.025,0.024)$&$(1.9604,1.9813)$&$333 \pm23 $&$23.02\pm0.21$ \\
$\eta'(\gamma\rho^0)\pi^-$           &$(-0.041,0.033)$&$(1.9618,1.9790)$&$1112\pm106$&$38.21\pm0.18$ \\
$\eta(\gamma\gamma)\rho^-$         &$(-0.058,0.041)$&$(1.9569,1.9855)$&$1801\pm113$&$24.43\pm0.10$ \\
\hline
SUM &  &  & $13157\pm240$ &  \\
\hline\hline
\end{tabular}
\end{table*}

\section{\boldmath Double Tagged $D_{s}^{+}$ events}
\label{Double Tagged}
\subsection{\bf Candidates for $D_{s}^{+}\rightarrow\eta(\eta')e^{+}\nu_{e}$}
Candidates for $D_{s}^{+}\rightarrow\eta(\eta')e^{+}\nu_{e}$
are selected on the recoil side of the ST $D_{s}^{-}$
and called as the double tagged (DT) event.
We require that (a) there is one charged track identified as an electron,
whose confidence level $CL_e$ is calculated by the $dE/dx$, TOF and EMC information
for the electron hypotheses, and satisfies $CL_e>0.001$ and $CL_e/(CL_e+CL_{\pi}+CL_K)>0.8$;
(b) the charge of the electron is opposite to the charge of the ST $D_{s}^{-}$ meson;
(c) $\eta (\eta')$ is reconstructed using the same criteria as those used in the ST $D^-_s$ selection;
(d) there is no extra charged track 
(and no extra $\pi^0$ for $D^+_s\to\eta' e^+\nu_e$) (Trk$_{\rm extra}$) 
except for those used in the DT event selection; 
(e) the maximum energy ($E_{\rm extra \gamma}^{\rm max}$) of the extra
photons, \emph{i.e.} those photons not used for reconstructing the
DT event, is required to be less than 300 MeV.

Due to the undetected neutrino, we cannot fully reconstruct the decay
$D_{s}^{+}\rightarrow\eta(\eta')e^{+}\nu_{e}$. However, we can
extract information on $D_{s}^{+}\rightarrow\eta(\eta')e^{+}\nu_{e}$ with the
missing energy and momentum in the event. To do so, we define a kinematic variable
$U_{\rm miss} \equiv E_{\rm miss}-|\overrightarrow p_{\rm miss}|$, where the missing energy $E_{\rm miss}$ and the missing
momentum $\overrightarrow p_{\rm miss}$ are calculated by the formulas
$E_{\rm miss} = E_{\rm cms} - \sum_j E_j$ and $\overrightarrow p_{\rm miss} = -\sum_j\overrightarrow p_j$,
in which $j$ runs over all the particles used to reconstruct the ST and DT candidates,
$E_j$ and $\overrightarrow p_j$
are the energy and momentum of the $j$th particle in the final state, and $E_{\rm cms}$
is the center-of-mass energy. Since only one neutrino is missing and
the neutrino mass is very close to zero, the $U_{\rm miss}$
distribution for signal events of
$D_{s}^{+}\rightarrow\eta(\eta')e^{+}\nu_{e}$ is expected to peak near zero.

\begin{figure*}[htbp]
\subfigure{\includegraphics[width=5.8cm, height=4.2cm]{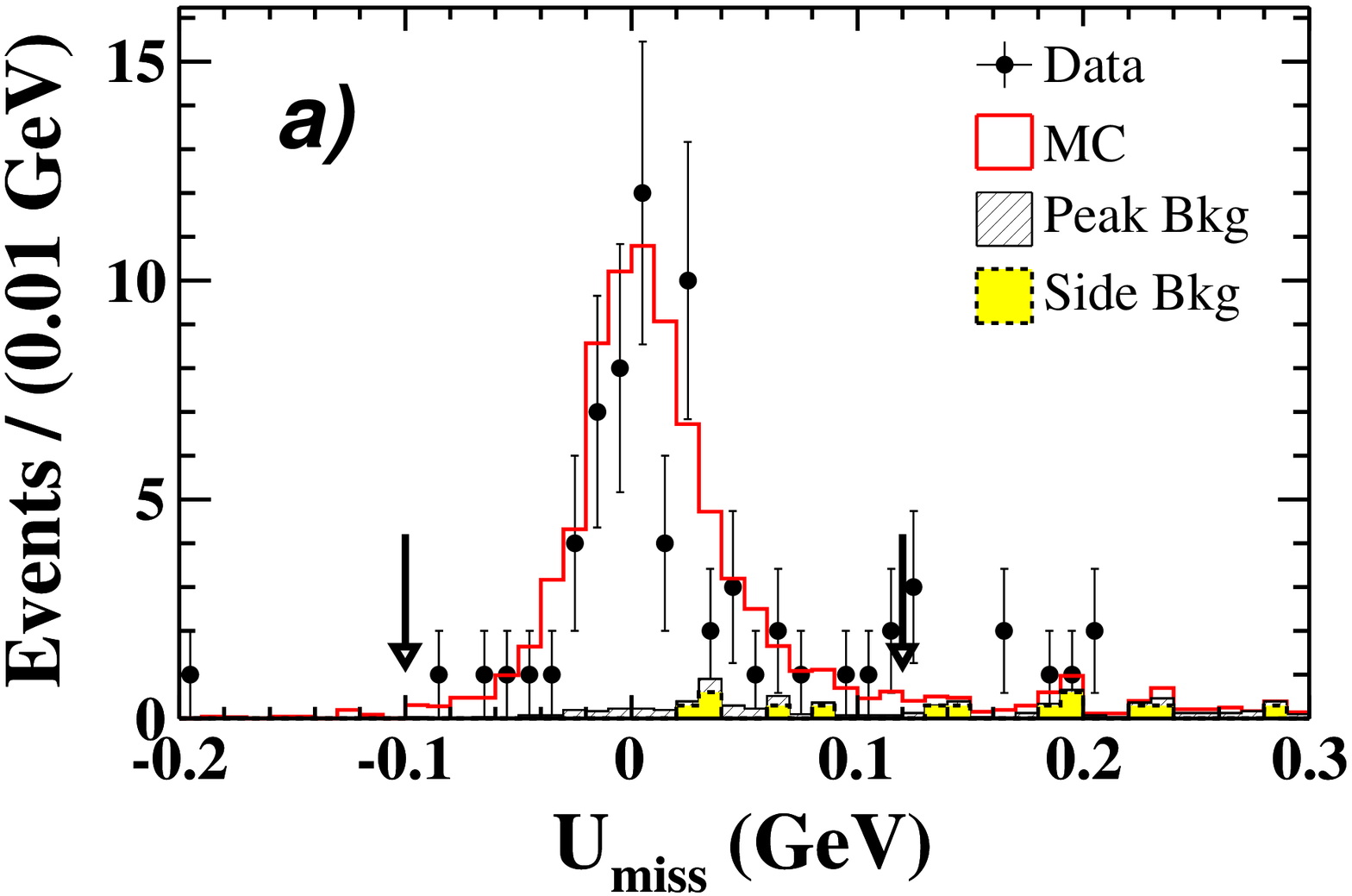}}
\subfigure{\includegraphics[width=5.8cm, height=4.2cm]{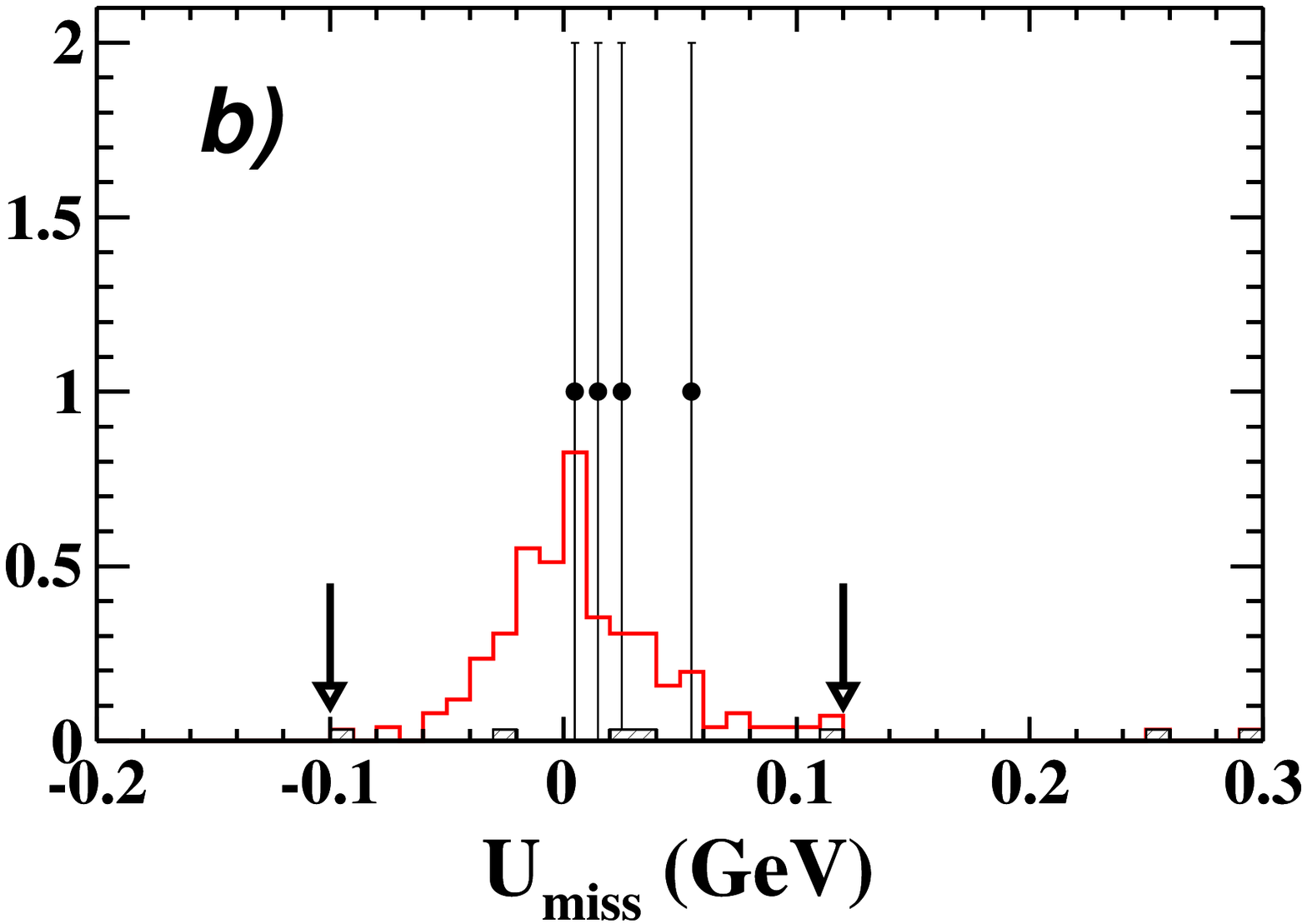}}
\subfigure{\includegraphics[width=5.8cm, height=4.2cm]{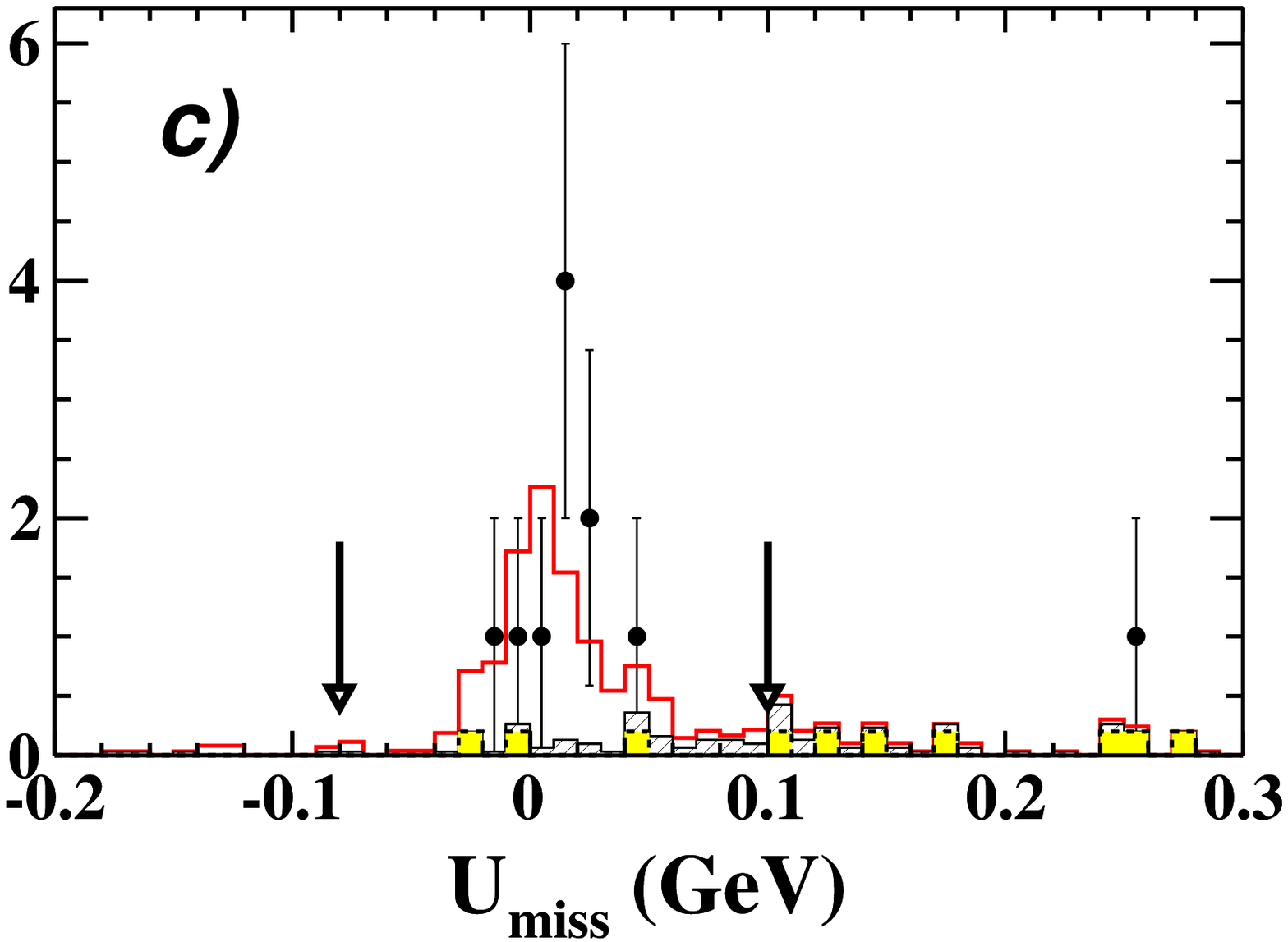}}
\caption{ \label{fig:Umiss_etaev_data_Inc}
Distributions of $U_{\rm miss}$ of the candidates for (a) $D_{s}^{+}\rightarrow\eta e^{+}\nu_{e}$,
(b) $D_{s}^{+}\rightarrow\eta'(\eta\pi^+\pi^-)e^{+}\nu_{e}$ and
(c) $D_{s}^{+}\rightarrow\eta'(\gamma\rho^0)e^{+}\nu_{e}$.
The pair of arrows indicates the signal region,
points with error bars show the events from data, the solid histograms show
the scaled events from inclusive MC, the hatched and
dashed histograms show the peaking background (`Peak Bkg') and
sideband backgrounds (`Side Bkg'), respectively.
}
\end{figure*}

Figure~\ref{fig:Umiss_etaev_data_Inc}
shows the $U_{\rm miss}$ distributions of the
candidates for $D_{s}^{+}\rightarrow\eta e^{+}\nu_{e}$,
$D_{s}^{+}\rightarrow\eta'(\eta\pi^+\pi^-)e^{+}\nu_{e}$, and
$D_{s}^{+}\rightarrow\eta'(\gamma\rho^0)e^{+}\nu_{e}$ in data.
The $U_{\rm miss}$ signal regions are defined as $(-0.10, 0.12)$~GeV, $(-0.10, 0.12)$~GeV and
$(-0.08, 0.10)$~GeV for $D_{s}^{+}\rightarrow\eta e^{+}\nu_{e}$,
$D_{s}^{+}\rightarrow\eta'(\eta\pi^+\pi^-)e^{+}\nu_{e}$ and
$D_{s}^{+}\rightarrow\eta'(\gamma\rho^0)e^{+}\nu_{e}$, respectively. Within the signal regions,
we observe $63.0\pm7.9$, $4.0\pm2.0$ and $10.0\pm3.2$ events, respectively.
\subsection{\bf Background estimate}
In the observed candidate events there are still some backgrounds,
which can be separated into two kinds.
The first kind is called the `peaking background' (Peak Bkg),
in which the ST $D_{s}^{-}$ is reconstructed correctly
and the semileptonic decay is reconstructed incorrectly.
To estimate this kind of background for $D_{s}^{+}\rightarrow\eta e^{+}\nu_{e}$,
we examine the inclusive $D_{s}^{+}D_{s}^{-}$ MC events with the signal events
excluded. After all selection criteria are applied, a total
of 82 events survive, which corresponds to an expectation of $2.6\pm0.3$ events for data.

The second kind is named the `sideband background' (Side Bkg),
in which the ST $D_{s}^{-}$ meson is reconstructed incorrectly.
This kind of background can be estimated by the events in the
$M_{\rm BC}$ sideband region, which is defined by the $M_{\rm BC}$ windows
of $(1.920, 1.950)$ and $(1.990, 2.000)$~GeV/c$^2$.
The number of backgrounds in the $M_{\rm BC}$
sideband region is then normalized
according to the background areas in signal and sideband region.
For $D_{s}^{+}\rightarrow\eta e^{+}\nu_{e}$,
$1.9\pm0.9$ `Side Bkg' events are observed.
Finally, we obtain the total number of background events to be $4.5\pm0.9$ for
$D_{s}^{+}\rightarrow\eta e^{+}\nu_{e}$.

For the decay $D_{s}^{+}\rightarrow\eta'e^{+}\nu_{e}$
with $\eta'\to\eta\pi^+\pi^-$ ($\gamma\rho^0$), the numbers of
`Peak Bkg' and `Side Bkg' events are estimated to be
$0.2\pm0.1$ ($1.2\pm0.2$) and $0.00^{+0.5}_{-0.0}$ ($0.6\pm0.4$), respectively.
The total numbers of the background events are $0.2^{+0.5}_{-0.1}$ and $1.8\pm0.4$
for $\eta'\to\eta\pi^+\pi^-$ and $\gamma\rho^0$ modes, respectively.

The $U_{\rm miss}$ distributions of the `Peak Bkg' and `Side Bkg' events for
$D^+_s\to\eta(\eta')e^+\nu_e$ are shown in Fig.~\ref{fig:Umiss_etaev_data_Inc}.

\subsection{\bf Net number of signals}
The numbers of observed candidate events and background events are summarized in Table~\ref{tab:Results_Data}.
After subtracting the numbers of background events, we obtain
the numbers of DT events ($N_{\rm DT}^{\rm net}$) to be
$58.5\pm8.0$, $3.8\pm2.0$ and $8.2\pm3.2$ for $D_{s}^{+}\rightarrow\eta
e^{+}\nu_{e}$, $D_{s}^{+}\rightarrow\eta'(\eta\pi^+\pi^-)e^{+}\nu_{e}$ and
$D_{s}^{+}\rightarrow\eta'(\gamma\rho^0)e^{+}\nu_{e}$, respectively.
\begin{table}[hbtp]
\begin{center}
\caption{\label{tab:Results_Data}
Observed event yields in data and expected background yields for
$D_{s}^{+}\rightarrow\eta e^{+}\nu_{e}$
and $D_{s}^{+}\rightarrow\eta'e^{+}\nu_{e}$.
}
\renewcommand\arraystretch{1.2}
\begin{tabular}{lccc} \hline\hline
Mode   & $N^{\rm obs}$ & $N^{\rm bkg}$ &$N_{\rm DT}^{\rm net}$  \\ \hline
$D_s^+\to\eta e^+\nu_e$&$63.0\pm7.9$&$4.5\pm0.9$&$58.5\pm8.0$ \\
$D_s^+\to\eta'(\eta\pi^+\pi^-)e^+\nu_e$&$4.0\pm2.0$&$0.2\pm0.1$&$3.8\pm2.0$  \\
$D_s^+\to\eta'(\gamma\rho^0)e^+\nu_e$  &$10.0\pm3.2$&$1.8\pm0.4$&$8.2\pm3.2$  \\ \hline\hline
\end{tabular}
\end{center}
\end{table}

\section{\boldmath Branching fractions}
The number of reconstructed ST $D_{s}^{-}$ events can be calculated from
\begin{equation}
N_{\rm ST} = 2\times
N_{D_{s}^{+}D_{s}^{-}}\times B_{\rm ST}\times\epsilon_{D_{s}^{-}}^{\rm ST},
\label{SingleTagEvts}
\end{equation}
where $N_{D_{s}^{+}D_{s}^{-}}$ is the number of
$D_{s}^{+}D_{s}^{-}$ meson pairs in data, $B_{\rm ST}$ is the branching
fraction for the ST $D_{s}^{-}$ decay,
$\epsilon_{D_{s}^{-}}^{\rm ST}$ is the ST efficiency. The number
of DT events for $D_{s}^{+}\to\eta(\eta')e^{+}\nu_{e}$
can be described as
\begin{equation}
\begin{aligned}
N_{\rm DT} & =  2\times N_{D_{s}^{+}D_{s}^{-}}\times B_{\rm ST} \\
& \times B(D_{s}^{+}\to\eta(\eta')e^{+}\nu_{e})
\times\epsilon^{\rm DT}_{D_{s}^{+}\to\eta(\eta')e^{+}\nu_{e}},
\label{DoubleTagEvts}
\end{aligned}
\end{equation}
\noindent where $B(D_{s}^{+}\to\eta(\eta')e^{+}\nu_{e})$ is the branching
fraction for $D_{s}^{+}\to\eta(\eta')e^{+}\nu_{e}$, and
$\epsilon^{\rm DT}_{D_{s}^{+}\to\eta(\eta')e^{+}\nu_{e}}$
is the efficiency of
simultaneously reconstructing the ST $D_{s}^{-}$ and
$D_{s}^{+}\to\eta(\eta')e^{+}\nu_{e}$
(DT efficiency). We can determine the
branching fraction for $D_{s}^{+}\to\eta(\eta')e^{+}\nu_{e}$ by
\begin{equation}
B(D_{s}^{+}\to\eta(\eta')e^{+}\nu_{e})=\frac{N_{\rm DT}^{\rm net}}{N_{\rm ST}^{\rm tot}\times
\epsilon_{D_{s}^{+}\to\eta(\eta')e^{+}\nu_{e}}\times B_i},
\label{BFSemlep}
\end{equation}
\noindent where
$\epsilon_{D_{s}^{+}\to\eta(\eta')e^{+}\nu_{e}} =
\epsilon^{\rm DT}_{D_{s}^{+}\to\eta(\eta')e^{+}\nu_{e}}/\epsilon^{\rm ST}_{D_{s}^{-}}$
is the efficiency of reconstructing
$D_{s}^{+}\to\eta(\eta')e^{+}\nu_{e}$, and
$B_i$ denotes the branching fractions for $\eta$ or $\eta'$ decays~\cite{pdg2014}.
The detection efficiencies are estimated using MC samples.
An simulated sample of $e^{+}e^{-}\to D_{s}^{+}D_{s}^{-}$ with
$D_{s}^{+}D_{s}^{-}$ decaying inclusively  is used to estimate the ST efficiency,
and a sample in which $D_{s}^{+}D_{s}^{-}$ decay exclusively into the ST modes accompanied by
$D_{s}^{+}\to\eta(\eta')e^{+}\nu_{e}$ is used
to estimate the DT efficiency.
The backgrounds associated with fake photon candidates, 
extra charged tracks and $\pi^0$
are correlated with the track multiplicity of the ST and signal modes. 
In this case, the requirements used to suppress these kinds 
of background events cause variations in the detection efficiencies
for $D_{s}^{+}\to\eta(\eta')e^{+}\nu_{e}$ 
between the different ST modes shown in Table~\ref{tab:Etaev_Eff}.
The detection efficiencies
for $D_{s}^{+}\to\eta(\eta')e^{+}\nu_{e}$
in the different ST modes are weighted by
the numbers of the ST
$D_{s}^{-}$ events; the average efficiencies are obtained to be ($49.04\pm0.21$)\%, ($16.16\pm0.13$)\% and
($24.20\pm0.16$)\% for $D_{s}^{+}\to\eta e^{+}\nu_{e}$,
$D_{s}^{+}\to\eta'(\eta\pi^+\pi^-)e^{+}\nu_{e}$ and
$D_{s}^{+}\to\eta'(\gamma\rho^0)e^{+}\nu_{e}$, respectively, as summarized in
Table~\ref{tab:Etaev_Eff}.

\begin{table*}[hbtp]
\begin{center}
\caption{\label{tab:Etaev_Eff}
Efficiencies $\epsilon_{D_{s}^{+}\to\eta(\eta') e^{+}\nu_{e}}
= \epsilon^{\rm DT}_{D_{s}^{+}\to\eta(\eta') e^{+}\nu_{e}}/\epsilon^{\rm ST}_{D_{s}^{-}}$ of reconstructing
$D_{s}^{+}\to \eta(\eta') e^{+}\nu_{e}$ in percentage, where
 $\epsilon^{\rm DT}_{D_{s}^{+}\to\eta(\eta') e^{+}\nu_{e}}$ and
$\epsilon^{\rm ST}_{D_{s}^{-}}$ are the DT and ST efficiencies which do not include the
branching fractions
$B(\pi^0\to\gamma\gamma)$, $B(K^{0}_{S}\to\pi^+\pi^-)$, $B(\eta\to\gamma\gamma)$,
$B(\eta'\to\eta\pi^+\pi^-)$ and $B(\eta'\to\gamma\rho^0)$.
The uncertainties are from MC statistics only.
}
\renewcommand\arraystretch{1.2}
\begin{tabular}{p{2.4cm} p{2.1cm}<{\centering} p{2.0cm}<{\centering} p{2.8cm}<{\centering} p{2.8cm}<{\centering} p{2.4cm}<{\centering} p{2.2cm}<{\centering}}\hline\hline
Tag Mode & $\epsilon^{\rm DT}_{D_{s}^{+}\to\eta e^{+}\nu_{e}}$ &
$\epsilon_{D_{s}^{+}\to\eta e^{+}\nu_{e}}$&
$\epsilon^{\rm DT}_{D_{s}^{+}\to\eta'(\eta\pi^+\pi^-) e^{+}\nu_{e}}$ &
$\epsilon_{D_{s}^{+}\to\eta'(\eta\pi^+\pi^-) e^{+}\nu_{e}}$ &
$\epsilon^{\rm DT}_{D_{s}^{+}\to\eta'(\gamma\rho^0) e^{+}\nu_{e}}$ &
$\epsilon_{D_{s}^{+}\to\eta'(\gamma\rho^0) e^{+}\nu_{e}}$ \\
\hline
$K^+K^-\pi^-$                       &$18.38\pm0.17 $&$47.22\pm0.45$&$5.79\pm0.10$&$14.89\pm0.27$&$8.72 \pm0.13$&$22.40\pm0.34$\\
$\phi(K^+K^-)\rho^-$                &$4.66 \pm0.07 $&$46.41\pm0.74$&$1.26\pm0.04$&$12.59\pm0.36$&$1.94 \pm0.04$&$19.30\pm0.46$\\
$K^{0}_{S}K^+\pi^-\pi^-$            &$10.71\pm0.14 $&$46.22\pm0.68$&$2.84\pm0.07$&$12.26\pm0.33$&$4.95 \pm0.10$&$21.36\pm0.44$\\
$K^{0}_{S}K^-\pi^+\pi^-$            &$10.32\pm0.14 $&$46.95\pm0.78$&$2.76\pm0.07$&$12.55\pm0.35$&$4.40 \pm0.09$&$20.04\pm0.46$\\
$K^{0}_{S}K^-$                      &$22.84\pm0.19 $&$50.80\pm0.48$&$7.85\pm0.12$&$17.46\pm0.28$&$11.81\pm0.14$&$26.27\pm0.33$\\
$\pi^+\pi^-\pi^-$                   &$25.58\pm0.20 $&$49.35\pm0.41$&$8.83\pm0.13$&$17.03\pm0.25$&$13.16\pm0.15$&$25.39\pm0.30$\\
$\eta(\gamma\gamma)\pi^-$           &$25.59\pm0.19 $&$53.78\pm0.48$&$9.85\pm0.13$&$20.71\pm0.30$&$13.75\pm0.15$&$28.90\pm0.35$\\
$\eta'(\eta\pi^+\pi^-)\pi^-$        &$11.43\pm0.14 $&$49.65\pm0.76$&$4.01\pm0.09$&$17.41\pm0.41$&$5.89 \pm0.21$&$25.58\pm0.95$\\
$\eta'(\gamma\rho^0)\pi^-$            &$19.18\pm0.18 $&$50.20\pm0.53$&$6.59\pm0.23$&$17.25\pm0.60$&$9.79 \pm0.13$&$25.62\pm0.37$\\
$\eta(\gamma\gamma)\rho^-$          &$12.68\pm0.15 $&$51.90\pm0.65$&$4.48\pm0.09$&$18.35\pm0.38$&$6.59 \pm0.11$&$26.99\pm0.47$\\
\hline
Weighted Average &--- & $49.04\pm0.21$ &--- & $16.16\pm0.13$ &--- & $24.20\pm0.16$\\ \hline\hline
\end{tabular}
\end{center}
\end{table*}

Inserting the numbers of $N_{\rm DT}^{\rm net}$, $N_{\rm ST}^{\rm tot}$, and
$\epsilon_{D_{s}^{+}\to\eta(\eta')e^{+}\nu_{e}}$
into Eq.~(\ref{BFSemlep}), we determine the branching fractions for
$D_{s}^{+}\rightarrow\eta e^{+}\nu_{e}$,
$D_{s}^{+}\rightarrow\eta'(\eta\pi^+\pi^-) e^{+}\nu_{e}$ and
$D_{s}^{+}\rightarrow\eta'(\gamma\rho^0) e^{+}\nu_{e}$
to be
$B(D_{s}^{+}\to\eta e^{+}\nu_{e)}= (2.30\pm0.31)\%$,
$B(D_{s}^{+}\to\eta'(\eta\pi^+\pi^-) e^{+}\nu_{e})= (1.07\pm0.56)\%$ and
$B(D_{s}^{+}\to\eta'(\gamma\rho^0) e^{+}\nu_{e})= (0.88\pm0.34)\%$, respectively.
To average the branching fraction for $D_{s}^{+}\rightarrow\eta'e^{+}\nu_{e}$,
we use a standard weighted least-squares procedure~\cite{pdg2014} and determine
it to be $B(D_{s}^{+}\to\eta' e^{+}\nu_{e})= (0.93\pm0.30)\%$.
With the measured branching fractions, we determine
the ratio to be
$\frac{B(D^+_s\to\eta' e^+\nu_e)}{B(D^+_s\to \eta e^+\nu_e)} = 0.40\pm0.14$,
where the uncertainties are statistical.

\section{\boldmath Systematic uncertainty}
In the measurement of the branching fractions for
$D_{s}^{+}\to\eta(\eta')e^{+}\nu_{e}$,
many uncertainties on the ST side mostly cancel in the efficiency
ratios in Eq.~(\ref{BFSemlep}).
Table~\ref{tab:sys_tot} summarizes the systematic uncertainties, which are discussed in detail below.
\begin{table}[hbtp]
\begin{center}
\caption{\label{tab:sys_tot}
Systematic uncertainties in percent in the measurements of the branching 
fractions for $D_{s}^{+}\rightarrow\eta e^{+}\nu_{e}$
and $D_{s}^{+}\rightarrow\eta'e^{+}\nu_{e}$. }
\renewcommand\arraystretch{1.2}
\begin{tabular}{p{2.8cm} p{1.0cm}<{\centering} p{2.5cm}<{\centering} p{1.6cm}<{\centering}}\hline\hline
Source &$\eta e^{+}\nu_{e}$ &
$\eta'(\eta\pi^+\pi^-)e^{+}\nu_{e}$ &
$\eta'(\gamma\rho^0)e^{+}\nu_{e}$ \\
\hline
Number of ST $D_{s}^{-}$                      & 1.8    & 1.8    & 1.8   \\
Tracking for $\pi^+$                            & ---    & 2.0    & 2.0   \\
PID for $\pi^+$                                 & ---    & 2.0    & 2.0   \\
Electron selection                            & 1.2    & 1.1    & 1.1   \\
$\eta(\eta')$ reconstruction                  & 2.3    & 2.5    & 2.8   \\
$E_{\rm extra \gamma}^{\rm max}$ cut          & 0.5    & 0.5    & 0.5   \\
Trk$_{\rm extra}$ veto                        & 0.4    & 1.4    & 1.4   \\
Background                                    & 0.5    & 0.7    & 0.8   \\
Weighted efficiency                           & 0.1    & 0.2    & 0.2   \\
Form factor model                             & 0.6    & 2.8    & 0.9   \\
MC statistics                                 & 0.4    & 0.8    & 0.7   \\
$B(\eta\to\gamma\gamma)$                      & 0.5    & 0.5    & ---   \\
$B(\eta'\to\eta\pi^+\pi^-)$                   & ---    & 1.6    & ---   \\
$B(\eta'\to\gamma\rho^0)$                     & ---    & ---    & 1.7   \\
$U_{\rm miss}$ requirement                    & 0.3    & 0.6    & 0.3   \\
\hline
Total                                         & 3.4    & 5.7    & 5.2   \\
\hline \hline
\end{tabular}
\end{center}
\end{table}

The uncertainty in the number of the ST $D_{s}^{-}$ mesons is estimated to be about 1.8\%
by comparing the difference between the fitted and the counted events in the
$M_{\rm BC}$ signal region.

The uncertainties in the tracking and PID for pion are both 1.0\% per track~\cite{Trk_pi}.
To investigate the uncertainty in the electron selection, we use Bhabha scattering events
as the control sample.
The efficiencies of the tracking and PID for electron
are weighted by
the polar angle and
momentum of the semileptonic decay.
The difference of efficiencies between data and MC is
assigned as the uncertainty in the tracking and PID for electron, which is
1.2\% (1.1\%) for $D^+_s\to\eta(\eta')e^+\nu_e$.

To estimate the uncertainty in the $\eta$ or $\eta'$ reconstruction,
including the uncertainty of photon detection efficiency, we analyze a control sample of
$\psi(3770)\to D^0\bar{D^0}$, where one $\bar{D^0}$ meson
is tagged by $\bar{D^0}\to K^+\pi^-$ or $\bar{D^0}\to K^+\pi^-\pi^-\pi^+$,
while another $D^0$ meson is reconstructed
in the decay $D^0\to K_S^0\eta$ or $D^0\to K_S^0\eta'(\eta'\to\pi^+\pi^-\eta$ or $\gamma\rho^0$).
The differences in the $\eta$ or $\eta'$ reconstruction
efficiencies between data and MC are estimated to be 2.3\%, 2.5\% and 2.8\%,
which are assigned as the uncertainties in the $\eta$ or $\eta'$ reconstruction
for $D_{s}^{+}\rightarrow\eta e^{+}\nu_{e}$,
$D_{s}^{+}\rightarrow\eta'(\eta\pi^+\pi^-)e^{+}\nu_{e}$ and
$D_{s}^{+}\rightarrow\eta'(\gamma\rho^0)e^{+}\nu_{e}$, respectively.

By examining the double tagged hadronic $D^*\bar{D}$ decays with
a control sample of $\psi(4040)\to D^*\bar{D}$, the
difference of the acceptance efficiencies with $E_{\rm extra \gamma}^{\rm max}< 300$~MeV
between data and MC is $(-0.18\pm0.33)$\%. We therefore assign 0.5\% as the uncertainty
in the $E_{\rm extra \gamma}^{\rm max}$ requirement.

The uncertainty due to the extra charged track and $\pi^0$ vetoes 
is estimated by analyzing
the fully reconstructed DT events
of $\psi(3770)\to D^+D^-$, where $D^-$ mesons are tagged by 
nine hadronic decay modes:
$K^+\pi^-\pi^{-}$,
$K^+K^-\pi^{-}$,
$K^{0}_{S}\pi^{-}$, $K^{0}_{S}K^{-}$, $K^{0}_{S}\pi^{+}\pi^{-}\pi^{-}$,
$K^{0}_{S}\pi^{-}\pi^{0}$, $K^+\pi^-\pi^{-}\pi^0$, 
$K^+\pi^-\pi^{-}\pi^-\pi^+$, $\pi^+\pi^-\pi^-$,
while $D^+$ mesons are reconstructed in the decay
$D^{+}\rightarrow\eta'\pi^+$. 
The data-MC difference 
in the reconstruction efficiencies 
with and without extra charged track and $\pi^0$ veto is 
assigned as the corresponding  systematic uncertainty,
which is estimated to be 0.4\% (1.4)\% for $D_s^+\to\eta(\eta') e^+ \nu_e$.

The uncertainty in the background estimate is determined by the uncertianties of
branching fractions~\cite{pdg2014} for the processes
$D_{s}^{+}\to\eta\mu^{+}\nu_{\mu}$, $D_{s}^{+}\to\rho^+\eta'(\eta\pi^+\pi^-)$
and $D_{s}^{+}\to\phi e^{+}\nu_{e}$, which are found to be the main background contributions
for $D_{s}^{+}\rightarrow\eta e^{+}\nu_{e}$,
$D_{s}^{+}\rightarrow\eta'(\eta\pi^+\pi^-)e^{+}\nu_{e}$ and
$D_{s}^{+}\rightarrow\eta'(\gamma\rho^0)e^{+}\nu_{e}$
from analyzing the MC sample. The systematic uncertainties are
estimated to be 0.5\%, 0.7\% and 0.8\%, respectively.

The uncertainty in the weighted efficiency estimate is mainly determined
by the weighting factors. Considering the statistical uncertainties of the weighting
factors in Table~\ref{tab:SingleTag}, we propagate them to
the uncertainty of the weighted efficiency during the calculation.
This uncertainty is estimated to be
0.1\% (0.2\%) for $D_{s}^{+}\rightarrow\eta(\eta') e^{+}\nu_{e}$.

The uncertainty in the form factor model of $D_{s}^{+}$ is determined by comparing the
detection efficiency to that with
a simple pole model (POLE,~\cite{POLE}). It is estimated to be
0.6\%, 2.8\% and 0.9\% for $D_{s}^{+}\rightarrow\eta
e^{+}\nu_{e}$, $D_{s}^{+}\rightarrow\eta'(\eta\pi^+\pi^-)e^{+}\nu_{e}$ and
$D_{s}^{+}\rightarrow\eta'(\gamma\rho^0)e^{+}\nu_{e}$, respectively.

The uncertainties in the MC statistics for
$D_{s}^{+}\rightarrow\eta e^{+}\nu_{e}$,
$D_{s}^{+}\rightarrow\eta'(\eta\pi^+\pi^-)e^{+}\nu_{e}$ and
$D_{s}^{+}\rightarrow\eta'(\gamma\rho^0)e^{+}\nu_{e}$,
which are determined by $\Delta\epsilon/\epsilon$,
where $\epsilon$ is the weighted average efficiency of reconstructing
$D_{s}^{+}\to \eta(\eta') e^{+}\nu_{e}$ and $\Delta\epsilon$ is the statistical uncertainty,
are 0.4\%, 0.8\% and 0.7\%, respectively.

The branching fractions for $\eta\to\gamma\gamma$, $\eta'\to\eta\pi^+\pi^-$
and $\eta'\to\gamma\rho^0$ are taken from PDG~\cite{pdg2014}.
Their uncertainties are 0.5\%, 1.6\% and 1.7\%, respectively.

To estimate the uncertainty in the $U_{\rm miss}$ requirement,
we examine the change in branching fractions when varying the $U_{\rm miss}$ signal region
by $\pm10$ or $\pm20$~MeV. The maximum changes of the branching
fractions are assigned as the uncertainties; they are found to be
0.3\%, 0.6\% and 0.3\% for
$D_{s}^{+}\rightarrow\eta e^{+}\nu_{e}$,
$D_{s}^{+}\rightarrow\eta'(\eta\pi^+\pi^-)e^{+}\nu_{e}$ and
$D_{s}^{+}\rightarrow\eta'(\gamma\rho^0)e^{+}\nu_{e}$,
respectively.

The total systematic uncertainties are obtained to be 3.4\%, 5.7\% and 5.2\%
for $D_{s}^{+}\rightarrow\eta e^{+}\nu_{e}$,
$D_{s}^{+}\rightarrow\eta'(\eta\pi^+\pi^-)e^{+}\nu_{e}$ and
$D_{s}^{+}\rightarrow\eta'(\gamma\rho^0)e^{+}\nu_{e}$, respectively,
by adding each of the uncertainties in quadrature.

In the measurement of $B(D_{s}^{+}\rightarrow\eta'(\eta\pi^+\pi^-)e^{+}\nu_{e})$
and $B(D_{s}^{+}\rightarrow\eta'(\gamma\rho^0)e^{+}\nu_{e})$,
the common systematic uncertainties are from the number of the ST $D_{s}^{-}$,
the tracking and PID for pion, electron selection, the $E_{\rm extra \gamma}^{\rm max}$ requirement,
extra tracks veto
and the weighted efficiency estimate.
The other systematic uncertainties are independent.  Finally, we assign 5.5\% as the
total systematic uncertainty for
$D_{s}^{+}\to\eta'e^{+}\nu_{e}$.

\section{\boldmath Summary}
\begin{table*}[htbp]
\caption{\label{tab:Comparison}
Comparison of the branching fractions
for $D_{s}^{+}\rightarrow\eta e^{+}\nu_{e}$ and
$D_{s}^{+}\rightarrow\eta'e^{+}\nu_{e}$ measured by BESIII Collaboration,
the previous measurements~\cite{CLEORatio,prd_80_052007,arx_1505_04205}
and the PDG values~\cite{pdg2014}.}
\begin{center}
\renewcommand\arraystretch{1.2}
\begin{tabular}{p{3.0cm} p{3.0cm}<{\centering} p{3.0cm}<{\centering} p{3.0cm}<{\centering} p{3.0cm}<{\centering} p{2.0cm}<{\centering}}\hline\hline
 &BESIII&Ref.~\cite{CLEORatio}& Ref.~\cite{prd_80_052007}&Ref.~\cite{arx_1505_04205}&PDG~\cite{pdg2014} \\ \hline
$B(D_{s}^{+}\rightarrow\eta e^{+}\nu_{e})$[\%]&
$2.30\pm0.31\pm0.08$& --- &$2.48\pm0.29\pm0.13$&$2.28\pm0.14\pm0.20$&$2.67\pm0.29$ \\
$B(D_{s}^{+}\rightarrow\eta'e^{+}\nu_{e})$[\%]&
$0.93\pm0.30\pm0.05$& --- &$0.91\pm0.33\pm0.05$&$0.68\pm0.15\pm0.06$&$0.99\pm0.23$ \\
$\frac{B(D_{s}^{+}\rightarrow\eta' e^{+}\nu_{e})}{B(D_{s}^{+}\rightarrow\eta e^{+}\nu_{e})}$
&$0.40\pm0.14\pm0.02$&$0.35\pm0.09\pm0.07$&---&---&--- \\ \hline \hline
\end{tabular}
\end{center}
\end{table*}
In summary, we measure the branching fractions for
$D_{s}^{+}\to\eta e^{+}\nu_{e}$ and $D_{s}^{+}\to\eta'e^{+}\nu_{e}$
to be $B(D_{s}^{+}\rightarrow\eta e^{+}\nu_{e})$ = ($2.30\pm0.31\pm0.08$)\%
and $B(D_{s}^{+}\rightarrow\eta' e^{+}\nu_{e})$ = ($0.93\pm0.30\pm0.05$)\%,
by analyzing the 482~pb$^{-1}$
data collected at $\sqrt s=4.009$~GeV with the BESIII
detector at the BEPCII collider with the double tag method,
and the ratio between $B(D_{s}^{+}\rightarrow\eta' e^{+}\nu_{e})$ and
$B(D_{s}^{+}\rightarrow\eta e^{+}\nu_{e})$ to be $0.40\pm0.14\pm0.02$,
where the first uncertainty is statistical and the second is systematic.
Table~\ref{tab:Comparison} shows a comparison of the branching fractions for
$D_{s}^{+}\rightarrow\eta e^{+}\nu_{e}$ and
$D_{s}^{+}\rightarrow\eta'e^{+}\nu_{e}$ as measured by
the BESIII Collaboration (this work),
previous measurements~\cite{CLEORatio,prd_80_052007,arx_1505_04205}
and the average values from PDG~\cite{pdg2014}.
The branching fractions measured in this work are in good
agreement with the previous measurements within uncertainties.
The ISGW2 model involves an $\eta-\eta'$ mixing angle close to
$-10^{\circ}$, which is the minimum value obtained from mass formulas~\cite{pdg2014} if
a quadratic approximation is used. According to Refs.~\cite{MixAngle,Glueball}, the measured ratio
is consistent with a pseudoscalar mixing angle of about $-18^{\circ}$.
Finally, the results improve upon the
$D_s^+$ semileptonic branching ratio precision
and provide more information for comprehensively
understanding the $D_s^+$ weak decays.

\section{\boldmath Acknowledgments}
The BESIII collaboration thanks the staff of BEPCII and the IHEP computing center for their strong support. This work is supported in part by National Key Basic Research Program of China under Contract No. 2015CB856700; National Natural Science Foundation of China (NSFC) under Contracts Nos. 11235011, 11322544, 11335008, 11425524; the Chinese Academy of Sciences (CAS) Large-Scale Scientific Facility Program; the CAS Center for Excellence in Particle Physics (CCEPP); the Collaborative Innovation Center for Particles and Interactions (CICPI); Joint Large-Scale Scientific Facility Funds of the NSFC and CAS under Contracts Nos. U1232201, U1332201; CAS under Contracts Nos. KJCX2-YW-N29, KJCX2-YW-N45; 100 Talents Program of CAS; National 1000 Talents Program of China; INPAC and Shanghai Key Laboratory for Particle Physics and Cosmology; German Research Foundation DFG under Contracts Nos. Collaborative Research Center CRC 1044, FOR 2359; Istituto Nazionale di Fisica Nucleare, Italy; Joint Large-Scale Scientific Facility Funds of the NSFC and CAS	 under Contract No. U1532257; Joint Large-Scale Scientific Facility Funds of the NSFC and CAS under Contract No. U1532258; Koninklijke Nederlandse Akademie van Wetenschappen (KNAW) under Contract No. 530-4CDP03; Ministry of Development of Turkey under Contract No. DPT2006K-120470; National Science and Technology fund; The Swedish Resarch Council; U. S. Department of Energy under Contracts Nos. DE-FG02-05ER41374, DE-SC-0010504, DE-SC0012069; U.S. National Science Foundation; University of Groningen (RuG) and the Helmholtzzentrum fuer Schwerionenforschung GmbH (GSI), Darmstadt; WCU Program of National Research Foundation of Korea under Contract No. R32-2008-000-10155-0.

\end{document}